# Dislocated accountabilities in the "AI supply chain": Modularity and developers' notions of responsibility

David Gray Widder[1]  and Dawn Nafus[2]

## Abstract
Responsible artificial intelligence guidelines ask engineers to consider how their systems might harm. However, contemporary artificial intelligence systems are built by composing many preexisting software modules that pass through many hands before becoming a finished product or service. How does this shape responsible artificial intelligence practice? In interviews with 27 artificial intelligence engineers across industry, open source, and academia, our participants often did not see the questions posed in responsible artificial intelligence guidelines to be within their agency, capability, or responsibility to address. We use Suchman's "located accountability" to show how responsible artificial intelligence labor is currently organized and to explore how it could be done differently. We identify cross-cutting social logics, like modularizability, scale, reputation, and customer orientation, that organize which responsible artificial intelligence actions do take place and which are relegated to low status staff or believed to be the work of the next or previous person in the imagined "supply chain." We argue that current responsible artificial intelligence interventions, like ethics checklists and guidelines that assume panoptical knowledge and control over systems, could be improved by taking a located accountability approach, recognizing where relations and obligations might intertwine inside and outside of this supply chain.

## Introduction and background

Many big technology companies are building responsible artificial intelligence (AI) programs[1] (Jobin et al., 2019), but those "owning" these programs are limited in their ability to create change, resulting in varying levels of efficacy (Metcalf et al., 2019). Even those without designated ethics roles are called to follow responsible AI guidelines (Jobin et al., 2019), checklists (Madaio et al., 2020), and other processes (Sirur et al., 2018). Outside of the biggest companies that build and deploy their own user-facing systems, many engineers operate at arm's length from their firm's immediate customer, who might themselves be multiple steps from a live deployment. How is responsibility and agency socially organized for AI practitioners in these distributed arrangements? What can be done in situations where responsibility is framed as checklist work and where this work risks falling through the cracks between actors?

We investigate how AI practitioners scope their agency and responsibility to address possible AI harms. Our participants described situations where they were asked to account for harms their systems may enable, yet saw those harms as beyond their agency, capability, or responsibility to address. We were struck by the deeply dislocated sense of accountability, where acknowledgement of harms was consistent, but nevertheless another person's job to address, always elsewhere. We suggest that the software engineering ideal of modularity, and the divisions of labor it enables, re-inscribe a belief in software production as supply chain, where developers recognize their dependence on others' code much like a shipment of goods: as necessary supplies, but not where a deep collaborative relationship might develop. When harms were recognized, it was usually through social locations cross-cutting or separate from the "supply chain." We argue that these same cross-

[1]School of Computer Science, Carnegie Mellon University, Pittsburgh, PA, USA
[2]Intel Labs, Portland, OR, USA

**Corresponding author:**
David Gray Widder, School of Computer Science,
Carnegie Mellon University, Pittsburgh, Pennsylvania, USA.
Email: david.widder@icloud.com

cutting locations can be used to rebuild responsible AI practice to recognize the limitations developers feel while building inter-organizational linkages that enable societal and commercial value.

Other work has shown that engineers do not see business relations within their scope to consider (Orr and Davis, 2020). Greene et al. (2019) showed that many responsible AI programs scrutinize AI system design instead of questioning the business purposes these systems enable. Familiar responsible AI interventions, like checklists, model cards, or data sheets, ask practitioners to map their technology to its end use, attempting to bring "out of scope" harms back in scope. We show how existing realities of software production work against this, catching developers between countervailing cultural forces.

The software engineering notion of "modularity" refers to a specific technical practice and the broader, inseparable cultural beliefs, epistemologies, and organizational arrangements it mediates and reinforces. Tech firms use metaphors of modular, containerized work to describe both code and teams of coders (Hanna and Park, 2020). Modularity has been a staple of software development since the 1970s, where large software systems are decomposed into smaller, self-contained parts, so one can control parts of a system without needing to address the myriad details of the other parts (Shaw, 2011). This "information hiding" (Parnas, 1972) buries "the complexity of each part behind an abstraction" (Baldwin et al., 2000: 64). This facilitates a division of labor and the matching of individual skills to specific tasks (Shaw, 2011) by separating concerns of different workers (Dijkstra, 1982). In practice, modular software may need fewer repairs and may be easier to repair, but software can also be *too* modular, perhaps due to error-prone and calcified inter-module interfaces (Kemerer, 1995). Nonetheless, open-source projects strive for modularity to make their codebase understandable (MacCormack et al., 2006), and professional software engineers see improved modularity as a benefit of refactoring their code (Kim et al., 2014).

This divided labor, inscribed in code itself, has enormous cultural and social implications. Modularity's apparent simplification facilitates the presence of "many hands" who are harder to keep accountable (Nissenbaum, 1996). The problem is more than many hands, however. Modularity sets the stage for a refusal to accept a relationship between "us" developers and "them" technology users, let alone other affected citizens (McPherson, 2018; Suchman, 2002). Others have noted that modularity is an epistemic culture (i.e. Cetina (1999)) that cultivates a capacity to "bracket off" (Malazita and Resetar, 2019), even when human beings are bracketed off, not pieces of code. This makes it an everyday form of the modernist fallacy of the separability of society from technology (Latour, 1993), separating code from harms it enables. It is an example of the social organization of ignorance (Proctor and Schiebinger, 2008), where the focus on one thing (the workings of a single portion of code) yields ignorance of another (the activities of other developers and users). This ignorance is not total, but situational: our participants were aware of harm, usually when outside of their role as a software engineer.

While other factors, including crude profit incentive, deepen this dislocated accountability, modularity is a touchstone of technical practice that serves as a lens through which these other matters are framed. Developers imagine their work as an extended series of modules that form a chain, as if the whole were a summation of parts. They also imagine that any particular piece of code is embedded in other code that is "near" or "far" to the general public (see Figure 1). By extension, entire organizations are also seen as "near" or "far" to end use, because organizations package up code to be "released" to other organizations. These are metaphors drawn from logistics. More than a metaphor, they also constitute the relations of logistics, from the obfuscation of distant labor practices to the security concerns that arise by not looking inside the "container" (Hockenberry, 2021).

Here, we focus on how the metaphor also defines other relations (business, personal reputations, user experience, etc.) as not part of the chain, but as a kind of secondary background. These "secondary" relations nevertheless hold things together in a different way. Carolan (2020), for example, follows Latour's (1999) "chains of translation," to examine data chains that tie the precision agriculture industry together in recursive and contested ways. While developers imagine supply chains as a series of upstream and downstream modules, like so many cargo containers awaiting shipment, Carolan's work suggests that chains can also work differently, where the links are not as discrete. Sociotechnical relations might occupy multiple social locations and cultural logics at the same time.

The links in a chain form a boundary of some kind, making responsibility "a boundary-crossing activity,

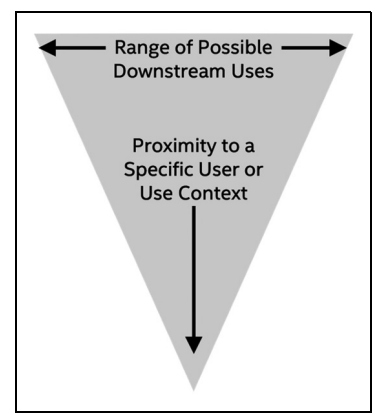

**Figure 1.** Work closer to a specific end-use context is perceived to imply a narrower range of possible (mis)uses.

taking place through the deliberate creation of situations that allow for the meeting of different partial knowledges" (Suchman, 2002: 94). We argue that asking developers to anticipate every conceivable outcome by diligently following elaborate checklists as if they occupied a view from nowhere (what Gansky and McDonald (2022) call "metadata maximalism") does not portend a meeting of partial knowledges. We take a located accountability approach that sees "systems development as entry into the networks of working relations" (Suchman, 2002: 92). In this context, that means asking developers to soften the view that once it is out of "my module"—the place that appears to make total knowledge possible—it is out of their control. Instead of metadata maximalism, we argue it is more effective to find and acknowledge where working relations can or do exist and where no single party has total knowledge or control. This is where developers can bring their partial, situated knowledge to bear. Even if technology use cannot be fully anticipated or controlled (Lally, 2021), crossing boundaries between "modules" can and does reduce ethical debt (i.e. Fiesler and Garrett, 2020). In this work, we identify key social locations that could create better points of boundary-crossing to reduce ethical debt. To conclude, we suggest that if accountability depends on the ability to critically analyze one's own social location, so that developers would have a better sense of to whom they are accountable and what they owe others in the chain, a thorny question arises. That is: what kind of critical reflection or questioning of modularity can be expected, given that modularity is itself a dominant form of social relations, and being located within social relations formed by modularity involves an injunction to reject the very notion of located accountability in the first place? We suggest three potential paths forward, depending on how deeply one is prepared to question modularity.

To conduct this study, we recruited using public emails and existing contacts, alongside paid services and snowball sampling to seek views from those working at various points in the AI supply chain, across different modalities of machine learning (ML) (i.e. computer vision and language processing) and application areas (i.e. military, manufacturing, and medicine). Our participants were not directly in the same supply chain such that we could trace a single component through it, but they did reflect patterns in what it meant to be "upstream" and "downstream." Our 27 participants were primarily in North America (16) and Europe (9), with one each in Asia and Africa. Private sector participants worked in eight companies ranging from startups to established smaller companies to large multinationals. Four researchers from three universities participated. Seven participants contributed to six open-source AI projects, sometimes as part of their employment, sometimes outside of it. Many had ML-related graduate degrees; job titles included Machine Learning Engineers, Research Scientists, Developer Experience Researchers, System Integrators, and Project Managers. All identified as men except one woman, reflecting disparities in the AI workforce. Each were invited to a semi-structured recorded teleconference interview, which were then professionally transcribed, except for one participant who preferred that we take notes. Most interviews lasted an hour, but were as short as 30 minutes or as long as two hours. After asking about their background, daily work tasks, and projects, we asked how they thought the system they are working on may be used or misused, where they saw possible harm, and if there was anything they wanted to, could, or currently do, to prevent it. A variety of interstitial documents accompanied our analysis. The first author wrote a descriptive memo after each interview including observations on how the participant described their agency on ethical questions, added to a running analytic memo documenting connections between participant accounts, and categorized quotes representing these connections iteratively. We also produced a table to reassemble the emic "supply chain" metaphor, which allowed us to identify patterns in how participants positioned their work on a spectrum from "general purpose" to "specific use." This became a resource for examining how the chain inflects views on responsibility. Our different positionalities helped us think critically about modularity, both from the standpoint of someone within computer science trained to see it as a valuable technical and social practice (Widder), and as someone trained to first see its epistemological shortcomings (Nafus).

In the next section, we illustrate how a distributed AI supply chain limits developers' sense of agency and responsibility. We then show the various ways the supply chain is reproduced in practice, alongside the social locations outside the chain that create space for responsible action to be taken. We show how the confluence of the two shapes the ethics work that is and is not done. Finally, we present three potential interventions, depending on one's view about whether modularity is an ideal to be preserved or a problem to be overcome.

## Views from up and down the AI supply chain

Outside of the largest technology companies, complex inter-organizational relationships are at the heart of building AI (Thomas, 2019). For example, computer vision used in a power plant's surveillance system to detect a person at its perimeter might begin life published as academic research, further developed and made freely accessible in an open-source library as a pretrained model, later requiring *in situ* training when deployed to work with the plant's existing hardware and software by a systems integrator. It might be further adapted if the plant has the requisite expertise.

Thomas (2019) observes that by 2018, computer vision professionals expected to not need to build systems from scratch, with open-source tooling and pretrained algorithms available to "kick start their work," and find a role somewhere in the chain. The persistence of a chain metaphor is notable given that software development professionals have shifted from linear "waterfall" production methods to nonlinear, iterative "agile" practices (Hockenberry, 2021; Gurses and van Hoboken, 2018). Chain metaphors come back into play precisely when developers imagine their scope of control, which they believe is limited by when a product is "released" by one organization and used by another. They also believe that control over their system's impacts increases as possible uses of the released system narrow, as it is adapted to fit a particular end use.

Higher in the AI supply chain are supposedly general purpose research outputs or tools, such as an academic ML researcher relaying his enjoyment in "discovering generalized infrastructure components that are missing from people's workflows, where "the application domain you pick can be potentially endless." This endlessness gives this person a sense of value and prestige, while the ability to control impacts does not. Separability between the optimization procedure and what is optimized sustains the belief that optimization tools are "general purpose," creating "endless" possibilities. From the top of the supply chain, the generality outshines the fact that there is a purpose of some kind, and that purpose precipitates some outcomes over others. Asked if there are ways his project could be used that would concern him, he answered: "nothing that would concern me [except] general ways in which you can abuse machine learning. […] I don't think it does anything that can be abused relative to what you could do normally with any machine learning algorithm." He extends the separation of the optimizing code from the optimized code to the people involved. This person at no point mentioned a "who" that might use his tool, suggesting he does not imagine there to be a social relation of some kind. He only imagines other inert containers of software, enabling him to normalize harmful ML practices as a general matter of course, or theoretical possibility, and not question his participation in it or his choices about who he allows to access his technology. His direct contribution to the "optimization" of harm by enabling it to occur in a more technically optimal manner is thus invisible. One might call it an uncritical technical practice (Agre, 1997), where incuriosity about the other person's "container" in turn leads to an incuriosity about why he is spending his time optimizing "the general ways you can abuse machine learning." Indeed, people working high in the supply chain were particularly prone to employ discourses of technological neutrality (i.e. Winner, 1980), referring to what they make as even more general purpose than the proverbial dual use gun: "I make a piece of equipment that makes pipe, somebody bought my pipe making equipment, and made the barrel of guns. I don't know how I stop [harm], because I didn't make the gun."

This view is also situated in a neoliberal economic context where *not* having relations or obligations is a dominant model of appropriate economic behavior (Grant, 1991; Callon, 1998). Unlike gift economies or other economic forms that constitute staples of economic anthropology (Plattner, 1989), the dominant narrative of economic exchange here is that there are no social ties after the exchange takes place. The parties are *quits*, with no further obligations to one another. This stands in stark contrast to the competing notions of responsible AI development found in the indigenous data sovereignty movement (Carroll et al., 2020), where care and the building of relations is central.

In the middle of the supply chain lie partial systems like performance benchmarks or pre-trained models, designed to show off accuracy, speed, or ease of use, as "kick starters" (Thomas, 2019) for others' future finished deployments. These contexts make upstream dependencies and downstream responsibilities more visible. For example, another engineer used "a composition of already existing components" from an open-source framework and models to develop machine translation "benchmarks," "showcase[s]," and "demo[s]," which he also made available as open-source. Because he did not build the framework, he stated "it's a part of open source project so […] we are not taking the full responsibility for the framework itself," downplaying whether he had any choice whether to vet it for problems. Looking downstream, he stated: "there is a very little interest in the actual…meaning of translation, but rather [more interest in] the performance numbers," like translation speed or accuracy. Because the output is not considered a final matter with real consequences, he does not consider it his job to address biases: "I don't believe that anyone will try to prove that, hey, the output is biased." While he was somewhat concerned that his company's logo would be attached, he expected the next person in the chain to know to address it, which re-rendered it as a "general" problem: "there is always a risk that the translation can be biased." He points to the least "general" actor in the chain as the site of responsibility: "I believe that the final responsibility lies at the client's side who is finally deploying the actual service." He frequently used passive voice to describe decisions that he could have made otherwise, for example, "the data was taken from official available sources" and "existing components, which are packed and prepared." These felt like statements of fact, not attempts to be exculpatory. The participant began the interview apologetically, explaining that his "very simple" project provided little for research on ethics.

Lower in the AI supply chain, an AI model is integrated into "live" software. Here, harms are closer and more visible, but managers still considered it a virtue for software engineers to be able to focus on their technical work, without interacting with those using their software. For example, a tech lead at a company building virtual reality services for defense industry clients explained that ethics are "a concern to me because there could be flaws in the code, security risks, quality risks, and effectively, if anything goes wrong, it looks bad on us." Nevertheless, he talks about the separation of engineers from colleagues that handle customer interactions with relief that he "kind of get[s] to turn a blind eye to certain social aspects" because "we have program managers that tend to be the buffer." He says sometimes he gets pulled into customer conversations but they are improving the process to make sure "I'm not involved, because frankly, I shouldn't be." If software engineers building the software might have an issue with their work being used to train military drone pilots, this separation insulates them from intimate knowledge of this use.

Downstream in the supply chain, the design affordances that limit use are more acknowledged. This participant was confident that his "app isn't so open ended that it can just be used […] by accident in a different way," noting it would take some reverse engineering to use it nefariously. But he is uncomfortable with his upstream dependencies, Facebook's Oculus: "we're kind of putting our foundation on sand" because "the platform […] is owned by Facebook [which[] recently had a pretty bad day [participant referencing then-recent congressional testimony]. So frankly, we don't trust them." This raises the real possibility that he may be vulnerable to having to pay his supplier's ethical debt (Fiesler and Garrett, 2020). Looking downstream, he is also aware of the care that needs to be taken with respect to which customers he does business with. He states, "There's always going to be some level of, let's say, customer qualification". Discussion of customer qualification did not occur higher in the supply chain.

When people talk in terms of "getting to turn a blind eye" to consequences and normalize harmful actions as a pervasive yet unconcerning matter, we have a form of social organization that creates a partial ignorance of customers and suppliers. To extend the logistics metaphor, these developers imagine themselves inside the container, not piloting the cargo ship or even developing the software that coordinates supply chain systems. Posner (2018) points out that in supply chains of physical goods, companies still struggle to gain full visibility into their networks of suppliers and labor or environmental conditions in part because the software that is supposed to create that transparency is as containerized as the goods and services it is meant to monitor. In this sense, the use of the supply chain metaphor is no coincidence; supply chains are a sociotechnical system of partial, selective sight (Posner, 2018). This "view from nowhere" (Haraway, 1991), then, is not a god's eye view, but a view from within a digital cargo container that knows little about where it heads. It is both difficult to know, because of the many hands problem, and there is little desire to know, because of the social organization of modularity. As Strathern (2002) reminds us, claims that technologies need to be set in some context already tell us about the context they are, in fact, in: one believed to lack social relations. In our case, modularity creates the numerous ways that responsibility is not to be found "here" regardless of where "here" is. Context is perennially displaced to elsewhere.

## Crosscurrents within and against the supply chain

In this section, we discuss ways that the supply chain is reproduced and the ways that people have to step out of the chain to prevent harm, whether in institutionally sanctioned or unsanctioned ways.

### Reproducing the supply chain

Divisions of labor, an important purpose of modularity, create the cracks through which responsible AI actions fall. It is remarkable that relationships themselves—acknowledging the effects that one person has on another—are seen by our participants as acts of labor that can be divided between people and handed off. This is neither a natural nor obviously normal state of affairs, as in other contexts the very notion of it would be utterly rejected (see Liboiron, 2021). In this context, however, to divide labor is so naturalized that participants expected relationships to either be rendered into a task or to not exist at all. One participant explained that no one tasked him with doing ethics work, so he doesn't do it: "I don't have time allocated during my normal week to think about […] responsible AI. This is not part of the work, at least not the part that someone would tell me from the top to worry about." There was often consternation about who would do an ethics assessment. A user experience researcher stated that ethics assessments are often filled out by software engineers and that "it was not my role" to do it. This posed a problem to him, because there "might be value in somebody who talks to customers i.e. me, filling it out versus an engineer," echoing work showing that separating concerns between UX and AI work is difficult (Subramonyam et al., 2022).

Status inflects divisions of labor. To the extent ethics was recognized enough to become a task, it was a task often seen as mere details. One participant filled out a privacy questionnaire for his team to use an existing data set to build a speech recognition benchmark. He felt the questionnaire asked for a lot of seemingly immaterial details his team was unconcerned with: "It wasn't that easy to get through all the sections [of the assessment…] there were some questions about how the storage is secured […] a team member of a research team or engineering team is not aware of [that] – it depends on IT support and configuration." Others simply handed the work off to contractors or junior employees, as a form of administrative labor no one else wanted to do. This is hardly a meeting of partial knowledge that would be suggested by taking located accountability seriously. Instead, it follows broader patterns of status between work on the model versus data (Sambasivan et al., 2021) and in programming generally (Coleman, 2012). Another university-based participant emphasized that he was encouraged to focus on results, which did not include the resulting societal impact of any kind: "It's not like when we're presenting [our research at a conference] they ask you […] what ethical steps did you take […] Usually they just want to see your result." These divisions made the authority to decide questions of ethics ambiguous. One participant building body scanning technology explained: "several questions [on the ethics assessment] are focused specifically on a machine learning AI statistical model, where many of the other questions are more around the broader product and business. So that was confusing," because making those assertions felt like an overstep of his own authority.

In addition to the division of labor, the pressure to "scale" to ever more data, users, and customers deepens the sense that others in the chain are unknowable and unconnected. For example, a tech lead for a virtual reality service (also referred to above) was concerned that while most of his current customers have been "physically met by one of our team at this point, that doesn't scale" as they build a service company. Another participant discussed a deep collaboration with a customer to build an AI system on the customer's site, but felt unable to know what the customer later did with that system, as follow-up work was believed to not scale, because it required labor to do it. Similarly, another participant said: "So right now, I know the clients. And we don't have clients [who do harmful things]. But in the future, once we go public you won't be even able to control that [… with] 10,000 clients – I don't know how many clients we'll get […] It can be difficult to track […] what they do with the system." His careful knowledge and consideration of his clients, the metaphorical glue between modules of the supply chain, is the very thing he also would have dismantled in his (and his company's) ideal future of broad adoption.

While these participants saw scale as a desirable state that creates a regrettable limitation on attention, others thought it legitimized not doing ethics work at all: "our company is so focused on growing and scaling with users that ethical AI is not really […] a big concern at this point." Others thought this would create friction and lose customers: "If you bring [ethical AI] for every other use case and every other customer, there is already a lot of customers that we are losing […] I don't want this to create a bottleneck for our customers" and be even a limitation on technological progress itself: "there is going to be hundreds of thousands of industrial uses of AI […] But if we start limiting ourselves from doing so because of ethical concern then it stops progress of so many developments."

No one in our study articulated a specific reason why one would want to scale; it was as if this was axiomatic enough to go unsaid. As Hanna and Park (2020) have argued, "scale thinking" is linked to modularity and capitalist impulses and is also its own perceived moral imperative that cannot be explained by economic or technical practices alone. These research participants are articulating the precise, embodied moments when scale becomes indifference: moments where conversation is severed, where the investment in care relationships wanes, and when context is no longer something one is a part of, moving from a situational awareness of harm (see also Madaio et al. (2022)) to a distant matter that needs "tracking." Participants invoked "scale" as a way of describing the removal of personal relations, as if it were impossible to know the motivations and desires of one's customers beyond individual personal connection, forgetting that there are entire business apparatuses designed to do so, like market research, customer management, or corporate auditing. What participants are expressing here is not a straightforward practical fact, but the way that notions of scale create a remoteness from reality that makes it possible to not see harm (Gray, 2021). Notions of scale render "technical systems as commodities that can be stabilized and cut loose from the sites of their production long enough to be exported *en masse* to the sites of their use" (Suchman, 2002: 95). They reinforce the distinction between inside and outside a company and create an important site of cutting a technology loose from its creators.

### Acting outside the supply chain

Social ties are not nearly as severed as the dominant discourse suggests. Participants were located in cultural logics that produce connections and responsible actions outside of the imagined triangle in Figure 1. Some of these activities are also the glue that holds the economic chains between organizations together, yet developers still saw themselves as stepping outside their supply chain role to act responsibly.

For example, being "customer-centric" was an explicit corporate value in many participants' workplaces that required them to understand how customers interact with their software to increase product satisfaction. User experience design plays a key role here. One participant led his team in a brainstorming session for their product to allow users to scan and monitor their body composition over time, which he felt was enabled by a shared and authentic "passion for the user, for the customer." To this end, they made design modifications in response to feedback from pilot studies with users, framing this as putting the customer's needs first: "We recognize [health and body composition as] a very sensitive thing […we…] focused on solving problems for the customer." While paying customers are often the privileged "humans" in "human"- centered design to the exclusion of other affected parties (Pasanen, 2019), specific anticipated users create a connection point between commercial incentives and better or worse societal impacts, even if these were proxies for relations rather than direct relations themselves.

User-centered design connects designer and engineer to (imagined or real) user, but mechanisms like licensing connect customer and supplier, especially further upstream. One participant's company released its ML framework both as freely available open source and as a download available only after signing up with an email address. Of these very different relationships, the participant preferred the second method because "we can be far more in touch with our customers. We know who they are, we can email them, we can make that more of a community." Being "in touch" clearly has economic value that notions of scale deny, but also holds potential to surface awareness of things that can go wrong downstream.

Marketing is another exchange point between actors. "Ethical AI" was seen as a marketing advantage, with one participant suggesting that it is a "very, very good influencing tool [where] users might choose [our company] over the competition." Another believed that responsible AI can be used to win sales: "the first thing that comes to mind is […] how to earn as much as possible, right? […] this Ethical and Responsible AI, [we are in a] world that using these terms could only help you, right?" Whether fortuitous alignment or crass co-opting, participants believed responsible AI efforts serve as a market differentiator, where companies can win business by helping their customers avoid ethical debt and the reputational costs it potentiates.

Similarly, engineers stepped out of the modules they build when thinking about how companies' ethical mishaps affect their own and their company's public reputation and profit. One participant relayed that his company had canceled a contract with a customer company which was using his team's software framework in a widely reported unethical way and suggested why this happened: a "public perception of your moral compass […] has a direct impact on your bottom line [which…] makes company owners stand up and do something different," namely, sever relations downstream. Participants directly associated with potentially harmful projects also feared personal reputational costs: "Some things can have uses that you don't intend, and that you don't want […] to come back to you." Concern about reputation seems the most direct acknowledgement of the impossibility of fully disconnected, modularized work. Developers know the impact of their creations will follow them or their companies when others believe it was their job to control the problem, even when they do not.

Reputation and customer value are not new frameworks for legitimizing ethics work (Metcalf et al., 2019). We should not interpret concern for reputation or attention to market value as always an indication of empty veneer. "Reputation" is the language through which social relations are acknowledged in a context that has an exceedingly thin vocabulary for them. Interviewees did not veer too far from their professional personas, where flat affect is the norm and private beliefs are expected to be contained into their own separate module. While we have little evidence, we suspect that for some, concern for reputation might reflect deeper notions of obligation for which there is no local vocabulary, while for others it might solely reflect concerns for economic consequences, while for others still, the two concepts might not be separate at all, and economic penalty might be taken as a sign of social disapproval. When participants wrestled with the problem of conflicted interests, the motives for reputational concern were questioned only when it came in the guise of other people. For example, one participant says he hears the term "ethical AI" from "C-suite kinds of people," but questioned whether this was a "buzzword" or whether something was "actually happening." While he believed his company doesn't want to "be a party to any inhumane usages of AI technologies" by downstream customers, he said they also want to "make money. And sometimes those are cross purposes." Similarly, another participant framed Google's treatment of Timnit Gebru as something that "communicates that they care about ethics [only] to a certain point."

There were also instances where corporate rationales were not what motivated ethical action. The participant working on the body scanning project, for instance, emphasized that his team's positive group dynamics was what made it possible to talk about ethics concerns by studying each other as pilot users, having their own bodies scanned, and sharing their intensely personal reactions. For this participant, ethics discussions were an exercise in vulnerability, and responsible design meant a powerful obligation of duty to one's colleagues and friends in the position of "user." While the technique has its limits (Bennett and Rosner, 2019), it is arguably more potent than hollow onstage rhetoric (i.e. Goffman, 1959) of "passion for the customer" or "human-centered design."

Ethics issues are not so easily disavowed when asked about work by friends and family: "It sometimes gets hard when other people ask me. […] 'What do you do?' […] 'Oh, I kind of - I work in the AI workspace?' 'Oh, so you're getting people killed and assassinated through - with drones […]' and it's like well, how much am I involved in that? […] You can't say it's not true because it is true. [AI] is used for that." Work on a "general purpose" framework did not allow him to unsee harms when called to account in social contexts. Others talked about wanting more from their employer. One person noted that they could not necessarily say whether their framework was being used by the US Army, and this not knowing was itself a kind of harm: "that's one thing I would really like to be informed, when my software is used. Where? For what purpose?"

We also heard of developers exerting a soft form of agency and resistance when their moral compass made them uncomfortable with assigned work (Wong, 2021). One participant's company's client asked them to track the actions of garment workers. Having inspected the training data the client provided, she stated, "It was a little sad looking at videos. They work from 6:00AM in the morning to 9:00PM at night." She said that even though the client called the project "object tracking," she was concerned that it would amount to algorithmic management: "the algorithm that we're using is basically looking at people's motions to figure out what exactly they are doing. So, sometimes […] they're just taking a break. You're just telling the system that this person's not doing anything." She described how her team deprioritized the project until the client pulled away: "[it was] not a project that any of us really wanted to work on. Thankfully it didn't go anywhere." This is softly subversive (Wong, 2021), in that subversion was undertaken through inaction rather than overt action. It is remarkable that otherwise elite and well-resourced AI developers nonetheless still feel they must resort to weapons of the weak (i.e. Scott, 1985). Whether caring for relations among coworkers or friends or for workers on a video who appear to be exploited, there is a quality of off-stage norm-making that is not encapsulated in official talk of "customer orientation" and responsible AI transparency interventions.

In practice, these crosscutting impulses to divide and connect lead to particular ways of handling responsibility and particular areas of priority. What does get attended to are matters of widespread public concern that can be encapsulated into a module of work without introducing friction into the development process. High-profile ethical lapses like racial and gender disparities in computer vision (Buolamwini and Gebru, 2018) and marquee regulatory action such as the European General Data Protection Regulation provide a shared social location, from outside the supply chain, from which to recognize some harms, but not others, within it. Bias might be measured statistically, but not questioned in other ways. For example, one participant doing AI research for the military was concerned about the mathematically identifiable biases within the weaponry, saying, "I think the whole issue of bias and its societal and ethical implications is terribly interesting and we don't have as much conversation, particularly with cyber weapons, as we should." Measurement fit the module, while any bias in the choices his customers might make about who to point weapons at did not.

This social configuration leaves us with an odd bimodality. On the one hand, prominent dramas about social harms embroil the careers of executives in congressional hearings, while on the other, contractors are asked to do "the paperwork." In a hollow middle, some limited actions do take place. Disparities in accuracy rates are often checked. Offstage action, like slowing work or meaningfully caring for a colleague playing the role of user, remains invisible, like a shadow responsible AI workforce with little connection to checklists, transparency, or customer vetting.

## Where to go from here?

Many efforts at supporting responsible AI, like AI fairness checklists (Madaio et al., 2020; Holstein et al., 2019), model cards (Mitchell et al., 2019), and datasheets (Gebru et al., 2021), assume panoptical visibility into the technology that our work demonstrates does not hold. Some have been designed as a kind of "nutrition label" (Chmielinski et al., 2022), where facts are announced to an unspecified audience as if taking a view from nowhere. Other toolkits, such as Vallor et al. (2018), acknowledge the interstitial nature of ethics failures, but when teams have neither visibility nor control over cascades of failure (i.e. Sambasivan et al., 2021), and do not believe they *should*, the success of inventory-like approaches is likely to be limited. If we instead start from an assumption of located accountability, where knowledge is partial and situated, we might seek places where there are relations between actors and where people who are not developers have a stronger role. While that is analytically straightforward in social scientific terms, it is more complicated for those who see the world through the lens of modularity and who value the cutting of relations for specific reasons we have shown. Therefore, asking others to simply adopt located accountability wholesale will not do. We see three possible approaches, depending on how much our colleagues trained in the virtue of modularity are willing to question it (Figure 2).

### Acting within the modules

If we fully accept that the dominance of modularity is unlikely to change soon, we would seek to act within it. Perhaps there is an opportunity for participants to *append* their partial understanding of the flaws, limitations, divergent provenances, and contexts of use of this documentation in checklists, model cards, and the like, thus relieving developers of the discomfort of being asked to definitively claim facts they felt they could not claim, as models and data changes hands.

This might require, ironically, doubling down on division of labor, by clearly delineating what knowledge on the card would come from developer's "module" and what comes from user experience, sales, and legal roles, leaving the supply chain metaphor largely intact. Nevertheless, this turns model cards into a boundary object where partiality comes together, even if deeper relations do not occur. This has obvious limitations. Unless there is a creative way to modularize participation from impacted groups, the very idea of which might be considered offensive, this approach re-inscribes their exclusion. It creates more modules for those who are not developers, but those include only what is publicly sayable. It is left to regulators, journalists, and academics to force conversation and action about that which is considered unsayable from within the chain.

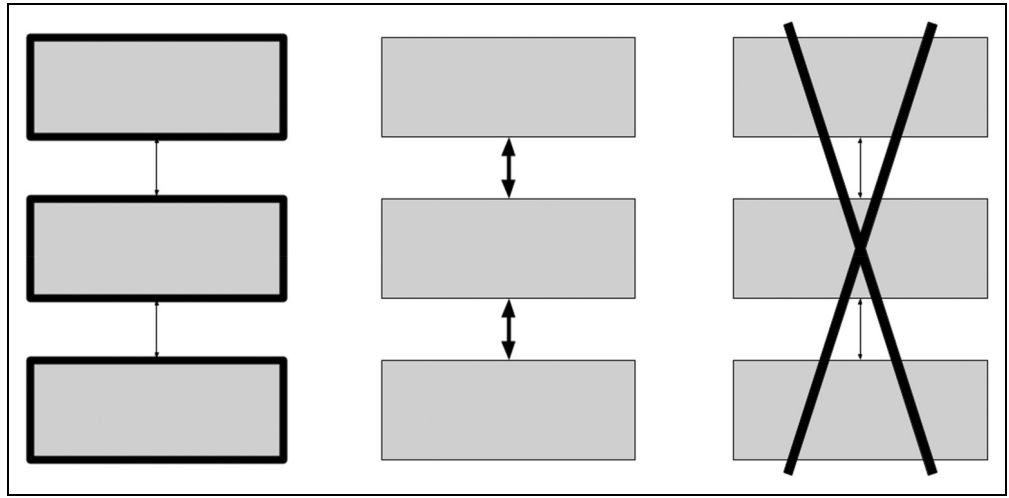

**Figure 2.** Three possible futures: (a) acting within the modules, (b) strengthening the interfaces, and (c) rejecting modularity.

### Strengthening the interfaces

Another approach would move away from metaphors of supply chains toward a more managerial notion of "value chains," which orchestrate companies' activities in ways that combine to create competitive advantages (Feller et al., 2006). This would strengthen business connections between companies beyond those allowed by the "developer hat" and buttress the communication that happens in the process of exchange. Model cards would be reinforced by contractual obligations and meaningful customer knowledge and communication, involving increased contribution from non-developers. Those in customer roles might scrutinize suppliers by asking for model cards, properly consented training data, and appropriate pay for data labelers, all scrutiny which is common in supply chains for physical goods. Javadi et al. (2021) propose technological measures strengthen module interfaces, by auditing AI services for misuse. These activities all help suppliers reframe ethics work as an act of delivering customer value. Still, this is not equally possible for every company. For example, one study showed some AI entrepreneurs concealed the ethics work they *were* doing from their venture capital funders, because they were interested in hiding limitations (Winecoff and Watkins, 2022).

The interface between onstage and offstage would have to be strengthened too, to help people integrate their multiple locations in and out of the supply chain. Developers might leverage their value as difficult-to-find laborers by making clear they are not prepared to pay personal reputational costs, while journalists and academics could also place more emphasis on the multi-actor cascades (Isaac and Hao, 2022). If the supply chain centers on perfect control over one's module, a value chain might center on probabilities and frictions—what technologies, contractual obligations, or marketing messages make easier or harder, faster, or slower. For example, the Ethical Source movement uses licenses to introduce legal friction for harmful uses in software supply chains, acknowledging this control is not total[2].

This approach facilitates the formation of stronger norms, bearing a surface relationship to Value Sensitive Design (Friedman, 1996). However, with numerous positionalities through the chain, "working misunderstandings" (Ferguson, 1994) in which parties mutually misrecognize the actions of one another, are more likely than straightforward values alignment. Managerial notions of "value chains" often elide the problem of *who* value is created for, on the assumption that value is a function of what markets will pay for. Depending on policy conditions, this approach could risk setting up a path dependence where ethics issues can be better acknowledged and acted upon, but remain a second-order, lagging concern where market value cannot be found.

### Rejecting modularity

What if modularity were eschewed entirely, both in terms of code and the broad social arrangements it mediates? Actors who object to the modularity ethos in the first place might abandon any notion of a chain entirely and prioritize building good relations as a matter of first order concern, building code second, and "scale" as a distant matter at most.

Here, social relations cannot be bracketed off as a mere input or requirements capture. The relationship is the objective, not the lines of code that may or may not result. Any code that does develop might be in the service of questioning what software tools are necessary at all and whether they need to be entirely different in different social conditions, as per Agre's critical technical practice (Agre, 1997). Echoing criticism of endless AI scale (Bender et al., 2021), Gebru and Hanna propose such a model of AI development, where the goal is not to produce "AI for the value of AI itself," but to instead be "sensitive to other forms of knowledge" in order to examine and curate data sets even if this is slower or more expensive (Strickland, 2022). Here, differences between users *do* make a difference (see Hanna and Park, 2020), while distinctions between producer and user begin to soften. One party is not the testbed for the other's "scale."

This approach might seem foreign to those building general purpose frameworks or scalable "software as a service" architectures. Look just outside dominant norms, however, and there are plenty of examples to be found. Indigenous data sovereignty principles specifically call for exactly this kind of approach (Carroll et al., 2020). In her work with North Carolina community healthcare workers building vaccine equity for Black and Latinx communities, Gray (2021) employed design justice principles from Costanza-Chock (2020) to argue that "we must prioritize a deep, methodical connection with subject matter and domain expertise in lieu of an unexamined rush to scale or to shield ourselves from the realities of a social world." Gray recognizes that her two-year intensive process introduced "friction, or working against scale, [which] is considered a bad thing in [Computer Science]. It is considered inefficient, a waste of engineering time." She recasts that ethos in the context of Arendt's banality of evil and notes that frictionless "efficiency" is the very thing that creates a remoteness from reality and opens the door to harm.

While the previous approach strengthened norms in a broad but inconsistent way, this does so in a more focused but deep way. Such focus has a long history outside an AI context (see Costanza-Chock, 2020, for an overview). However, rejecting modularity in a modularized world raises interesting questions for upstream tools. Would someone fully reject all lines of code that were ever designed as modules in a literal way, or reject the broader belief system modularity entails and seek opportunities to build differently, or be more careful about choices of upstream components, like libraries or compilers, especially when made by companies known for ethics breaches? These choices might open up new avenues of technical innovation. In making them, teams might learn the specific ways that "generic" tools are not in fact generic at all, but generic only to those who are currently well served by the current supply chain. It might be that the need for other kinds of yet undeveloped "generic" tools that serve other interests becomes apparent. Finding and developing these would be a significant act of critical technical practice and open up engineering paths otherwise foreclosed. This approach also raises questions for public policy. Given the resource inequalities between community groups and companies that seek to scale, and that those same groups are meeting social needs that arguably benefit a country as a whole, what would an appropriate science and technology policy do to support these efforts?

## Conclusion

Thinking about ethics and responsibility as chains of relations reveals specific locations in which ethical decision-making can take place. Those locations might be upstream or down, and they might be within the cultural logic of modularity or outside it. The combinations of these locations shape what is considered sayable and what is off-stage talk. They shape what is prestige-garnering work, what is paperwork, and what is high stakes public drama. These social locations also shape the points of AI governance intervention, which rely on the extent to which actors themselves are willing to, and are capable of, acknowledging their own locations within a broader system of production and engaging more fully in the relations in which they are involved. The core of the matter—how much modularized thinking should dominate software production—will not be settled easily. Consensus might not be achieved, and multiple paths might be followed by different sets of actors with different visions of what responsibility is. Regardless of which directions others take, we have shown that realistic responsible AI interventions can start by making deliberate choices about how strong a role current software production ideals should play in future responsible AI development.

## Acknowledgements

Whereas this paper is a module of work, the ideas in it are the product of a web of ties to Karly Burch, Laura Dabbish, Henry Fraser, Jim Herbsleb, Michael Madaio, Lama Nachman, John Sherry, and Julie Widder, whom we gratefully thank for their help. This work was conducted with support from Intel Corporation. We dedicate this work to Suzanne Thomas, who encouraged this work in its early stages and whose scholarship and partnership touched us and many others.

## Declaration of conflicting interests

The author(s) declared no potential conflicts of interest with respect to the research, authorship, and/or publication of this article.

## Funding

The author(s) disclosed receipt of the following financial support for the research, authorship, and/or publication of this article: This work was supported by the Intel Corporation.

## ORCID iDs

David Gray Widder  https://orcid.org/0000-0002-6912-8067
Dawn Nafus  https://orcid.org/0000-0002-2976-6662

## Notes

1. "Responsible AI" as opposed to "ethical AI" appears to be the more common term. Our own use of "responsible AI" denotes our commitment to feminist theories of technology (Haraway, 1991), where ethics cannot be removed from the question of "to whom?" does one owe a response. We sometime use "ethical AI" where context makes it appropriate.
2. See: https://ethicalsource.dev.

12	Big Data & SocietyLatour B (1993) *We Have Never Been Modern*. Cambridge, MA: Harvard University Press.

Latour B (1999) *Pandora's Hope: Essays on the Reality of Science Studies*. Cambridge, MA: Harvard University Press.

Liboiron M (2021) *Pollution Is Colonialism*. Durham, NC: Duke University Press.

MacCormack A, Rusnak J and Baldwin CY (2006) Exploring the structure of complex software designs: An empirical study of open source and proprietary code. *Management Science* 52(7): 1015–1030.

Madaio M, Egede L, Subramonyam H, et al. (2022) Assessing the fairness of AI systems: AI practitioners' processes, challenges, and needs for support. ACM Conference on Computer-Supported Cooperative Work and Social Computing 6, CSCW1 (2022): 1–26.

Madaio MA, Stark L, Vaughan JW, et al. (2020) Codesigning checklists to understand organizational challenges and opportunities around fairness in AI. In ACM Conference on Human Factors in Computing Systems: 1–14.

Malazita JW and Resetar K (2019) Infrastructures of abstraction: How computer science education produces anti-political subjects. *Digital Creativity* 30(4): 300–312.

McPherson T (2018) *Feminist in a Software Lab: Difference + Design*. 6. Cambridge, MA: Harvard University Press.

Metcalf J, Moss E, et al. (2019) Owning ethics: Corporate logics, silicon valley, and the institutionalization of ethics. *Social Research: An International Quarterly* 86(2): 449–476.

Mitchell M, Wu S, Zaldivar A, et al. (2019) Model cards for model reporting. In ACM Conference on Fairness, Accountability, and Transparency: 220–229.

Nissenbaum H (1996) Accountability in a computerized society. *Science and Engineering Ethics* 2: 25–42. https://doi.org/10.1007/BF02639315.

Orr W and Davis JL (2020) Attributions of ethical responsibility by artificial intelligence practitioners. *Information, Communication & Society* 23(5): 719–735.

Parnas David Lorge (1972) On the criteria to be used in decomposing systems into modules. *Communications of the ACM*: 1053–1058.

Pasanen J (2019) Human-centred design considered harmful. (2019). Accessed: 2022-05-2.

Plattner S (1989) *Economic Anthropology*. Redwood City, CA: Stanford University Press.

Posner M (2018) See no evil. Logic Magazine, Issue 4: "Scale".

Proctor RN and Schiebinger L (2008) Agnotology: The making and unmaking of ignorance. (2008).

Sambasivan N, Kapania S, Highfill H, et al. (2021) "Everyone wants to do the model work, not the data work": Data cascades in high-stakes AI. In ACM Conference on Human Factors in Computing Systems: 1–15.

Scott and James (1985) *Weapons of the Weak: Everyday Forms of Peasant Resistance*. New Haven, CT: Yale University Press.

Shaw M (2011) Modularity for the modern world: Summary of invited keynote. In international conference on aspect-oriented software development. 1–6.

Sirur S, Nurse JR and Webb H (2018) Are we there yet? Understanding the challenges faced in complying with the General Data Protection Regulation (GDPR). In 2nd international workshop on multimedia privacy and security: 88–95.

Strathern M (2002) Abstraction and Decontextualization: An anthropological comment. In: Woolgar Steeve (ed) *Virtual Society?: Technology, Cyberbole, Reality*. Oxford, United Kingdom: Oxford University Press, 302.

Strickland E (2022) Timnit Gebru is building a slow AI movement. https://spectrum.ieee.org/timnit-gebru-dair-ai-ethics. *IEEESpectrum* (31 March 2022). Accessed: 2022-07-7.

Subramonyam H, Im J, Seifert C, et al. (2022) Solving separation-of-concerns problems in collaborative design of human-AI systems through leaky abstractions. In ACM Conference on Human Factors in Computing Systems: 1–21. Association for Computing Machinery.

Suchman L (2002) Located accountabilities in technology production. *Scandinavian Journal of Information Systems* 14(2): 7.

Thomas SL (2019) Migration versus management: The global distribution of computer vision engineering work. In 2019 ACM/IEEE 14th International Conference on Global Software Engineering. IEEE, 12–17.

Vallor S, Green B and Raicu I (2018) Ethics in technology practice: A Toolkit. The Markkula Center for Applied Ethics at Santa Clara University. https://www.scu.edu/ethics-in-technology-practice/ethical-toolkit/ (2018).

Winecoff AA and Watkins EA (2022) Artificial concepts of artificial intelligence: Institutional compliance and resistance in AI startups. arXiv preprint arXiv:2203.01157 (2022).

Winner L (1980) Do artifacts have politics? *Daedalus*. 109: 121–136.

Wong RY (2021) Tactics of soft resistance in user experience professionals' values work. Proceedings of the ACM on Human-Computer Interaction, 5, CSCW2 (2021), 1–28.